\documentclass[twocolumn,showpacs,preprintnumbers,amsmath,amssymb]{revtex4}
\usepackage{graphicx}% Include figure files
\usepackage{dcolumn}% Align table columns on decimal point
\usepackage{bm}% bold math
\usepackage{longtable}
\def\bea {\begin{eqnarray}}
\def\eea {\end{eqnarray}}

\def\be {\begin{equation}}
\def\ee {\end{equation}}

\begin{document}
\title{High spin band structures in doubly-odd $^{194}$Tl }
\author{H. Pai} 
\author{G. Mukherjee}
\thanks{Corresponding author}
\email{gopal@vecc.gov.in}
\author{S. Bhattacharyya}
\author{M.R. Gohil}
\author{T. Bhattacharjee}
\author{C. Bhattacharya}
\affiliation{Variable Energy Cyclotron Centre, Kolkata 700064, India}
\author{R. Palit, S. Saha, J. Sethi, T. Trivedi, Shital Thakur, B.S. Naidu, S.K. Jadav, R. Donthi}
\affiliation{Department of Nuclear and Atomic Physics, Tata Institute of Fundamental Research, Mumbai-400005, India}
\author{A. Goswami}
\affiliation{Saha Institute of Nuclear Physics, Kolkata 700064, India}
\author{S. Chanda}
\affiliation{Fakir Chand College, Diamond Harbour, West Bengal, India }
\date{\today}
\begin{abstract}
The high-spin states in odd-odd $^{194}$Tl nucleus have been studied by populating them using the 
$^{185,187}$Re($^{13}$C, xn) reactions at 75 MeV of beam energy. A $\gamma-\gamma$ coincidence measurement 
has been performed using the INGA array with a digital data acquisition system to record the time stamped 
data. Definite spin-parity assignment of the levels was made from the DCO ratio and the IPDCO ratio 
measurements. The level scheme of $^{194}$Tl has been extended up to 4.1 MeV in excitation energy including 
19 new $\gamma$-ray transitions. The $\pi h_{9/2} \otimes \nu i_{13/2}$ band, in the neighboring odd-odd Tl
isotopes show very similar properties in both experimental observables and calculated shapes. Two new band 
structures, with 6-quasiparticle configuration, have been observed for the first time in $^{194}$Tl. One of 
these bands has the characteristics of a magnetic rotational band. Cranked shell model calculations, using 
a deformed Woods-Saxon potential, have been performed to obtain the total Routhian surfaces in order to study 
the shapes of the bands and the band crossing in $^{194}$Tl. The semiclassical formalism has been used to describe 
the magnetic rotational band.

\end{abstract}

\maketitle
\section{Introduction}
The Thallium nuclei, with proton number Z = 81, are situated in a transition region between the deformed prolate 
rare earth nuclei and the spherical lead nuclei at Z = 82. The proton Fermi level in Tl lies below the Z = 82 shell
closure and near the 2s$_{1/2}$ orbtal. The ground state spin-parity of the heavier odd-A Thalium isotopes, are 
accordingly, 1/2$^+$ \cite{odd-Tl1,odd-Tl2,odd-Tl3}. However, the intruder 9/2$^-$[505] and 1/2$^-$[541] Nilsson 
states, originating from the $\pi$h$_{9/2}$ orbital above Z = 82, are also available near the Fermi surface for 
oblate and prolate deformations, respectively. The experimental evidence comes from the observation of a low 
lying 9/2$^-$ isomeric state and strongly coupled rotational bands built on this in the odd-A Thalium isotopes 
\cite{odd-Tl4,1,2,3,4}. The neutron Fermi surface, for the Tl and Hg isotopes in the $A\sim 190$ region lie near the 
top of the $1i_{13/2}$ orbital and decoupled bands with this configuration have been identified in odd-A Hg 
(Z = 80) isotopes in this region \cite{5,6}. Therefore, in even-mass Tl isotopes in this region, collective 
rotational bands based on the $\pi h_{9/2} \otimes \nu i_{13/2}$ configuration are expected. Such bands have been 
experimentally observed in a few odd-odd Tl isotopes \cite{7,8,9,10,11,12,13}. But in most of the cases, there 
are ambiguities on level energies, spins and parities. Recently, experimental studies on the structures of 
$^{190}$Tl \cite{7} and $^{198}$Tl \cite{13} have been reported with definite spin-parity assignment of the 
$\pi h_{9/2} \otimes \nu i_{13/2}$ band. In $^{190}$Tl, this band, which shows a low spin signature inversion, 
was reported to have an oblate structure. Whereas in $^{198}$Tl, the possible chiral structure associated with 
this band was interpreted with a triaxial deformation. Therefore, in order to understand the possible
transition from oblate to triaxial shape induced by the change in the neutron Fermi surface of Tl nuclei, it is 
important to chracterize the $\pi h_{9/2} \otimes \nu i_{13/2}$ band systematically in an isotopic chain. But 
the high spin data on the other odd-odd Tl nuclei are scarce. Moreover, both the protons and the neutrons occupy 
high-j orbitals so, different kinds of collective and single particle excitations, like magnetic rotation, are 
expected. Although several magnetic rotational bands have been observed in Pb and Hg nuclei in this region \cite{14, 15}, 
no such bands are reported for Tl isotopes.

The knowledge of high spin states in $^{194}$Tl was very limited prior to the present work although several superdeformed
bands have been reported for this nucleus by F. Azaiez et al., but these bands were not connected with the normal deformed 
bands \cite{aza90, aza91}. The normal deformed high spin states in $^{194}$Tl were studied by Kreiner {\it et al.} \cite{9} 
and a level scheme was obtained using two Ge(Li) detectors. Although indication of rotatonal bands based on the $\pi h_{9/2} 
\otimes \nu i_{13/2}$ configuration was reported in that work, no definite spin-parity assignments could be done. Possible 
observation of chiral partner bands in $^{194}$Tl has been claimed recently \cite{194Tl2009} but no such level scheme has been 
reported till date. In this work, we have studied the $\gamma$-ray spectroscopy of $^{194}$Tl with the aim, in particular, to 
compare the band structures in odd-odd Tl isotopes and to search for a possible magnetic rotational band. 

\section{Experimental Method and Data Analysis}

The $\gamma$-ray spectroscopy of $^{194}$Tl has been studied at 14-UD BARC-TIFR Pelletron at Mumbai, India using the 
Indian National Gamma Array (INGA). The INGA consisted of 15 clover HPGe detectors with BGO anti-Compton shields 
at the time of the experiment. The excited states of $^{194}$Tl were populated by fusion evaporation reactions 
$^{185,187}$Re($^{13}$C, xn)$^{194}$Tl at the beam energy of 75 MeV. The target was a thick (18.5 mg/cm$^2$) 
natural rhenium target. The recoils were stopped inside the target. The isotopic ratio of $^{185}$Re and 
$^{187}$Re in the natural rhenium is 37:63. According to the PACE-IV calculations, the 4n and 5n channels are 
the dominant ones at the beam energies encompassed inside the target in this experiment. The beam energy at the 
exit of the target was calculated to be below the coulomb barrier. The clover detectors were arranged in six 
angles with 2 clovers each at $\pm$40$^\circ$ and $\pm$65$^\circ$ while four clovers were at 90$^\circ$ and 
three were at -23$^\circ$ angles. The average count rate in each crystal was limited to $\sim$ 4k/sec. 
The clover detectors were calibrated for $\gamma$ ray energies and efficiencies by using $^{133}$Ba and 
$^{152}$Eu radioactive sources. 

Recently, a digital data acquisition (DDAQ) system, based on Pixie-16 modules developed by XIA LLC \cite{ta08}, 
has been adopted for the INGA. This system has provision for the digitization of 96 channels 
of 24 clover detectors (maximum number of clover detectors that can be put in INGA) with 100 MHz sampling 
rate. This DDAQ system was used in the present experiment for the data collection. Time stamped data were 
collected when at least two clovers (Compton suppressed) were fired in coincidence. A time window of 150 ns 
was set for this coincidence between the fast triggers of individual channels and the coincidence trigger 
was kept open for 1.5 $\mu$s. The BGO signals from the anti-Compton shields of the respective clovers 
were used for vetoing the individual channels. The detailed description of the DDAQ has been given in the 
Refs. \cite{dsp1, dsp2}. 

%%%%%%%%%%%%%%%%%%%%%%%%%%%%%%%%%%%%%%%%%%%%%%%%%%%%%%%%%%%%
\begin{figure}[ht]
\begin{center}
\includegraphics*[scale=0.31, angle = 0]{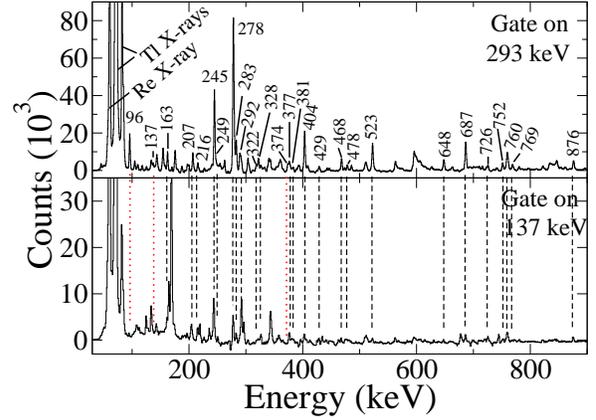}
\caption{(Color online) Coincidence spectra by gating on 293-keV (top) and 137-keV (bottom) $\gamma$ transitions. 
The unmarked peaks are the contaminants.}
\label{fig1}
\end{center}
\end{figure}
%%%%%%%%%%%%%%%%%%%%%%%%%%%%%%%%%%%%%%%%%%%%%%%%%%%%%%%%%%%%

%%%%%%%%%%%%%%%%%%%%%%%%%%%%%%%%%%%%%%%%%%%%%%%%%%%%%%%%%%%%
\begin{figure*}[ht]
\begin{center}
\includegraphics*[scale=0.6, angle = 0]{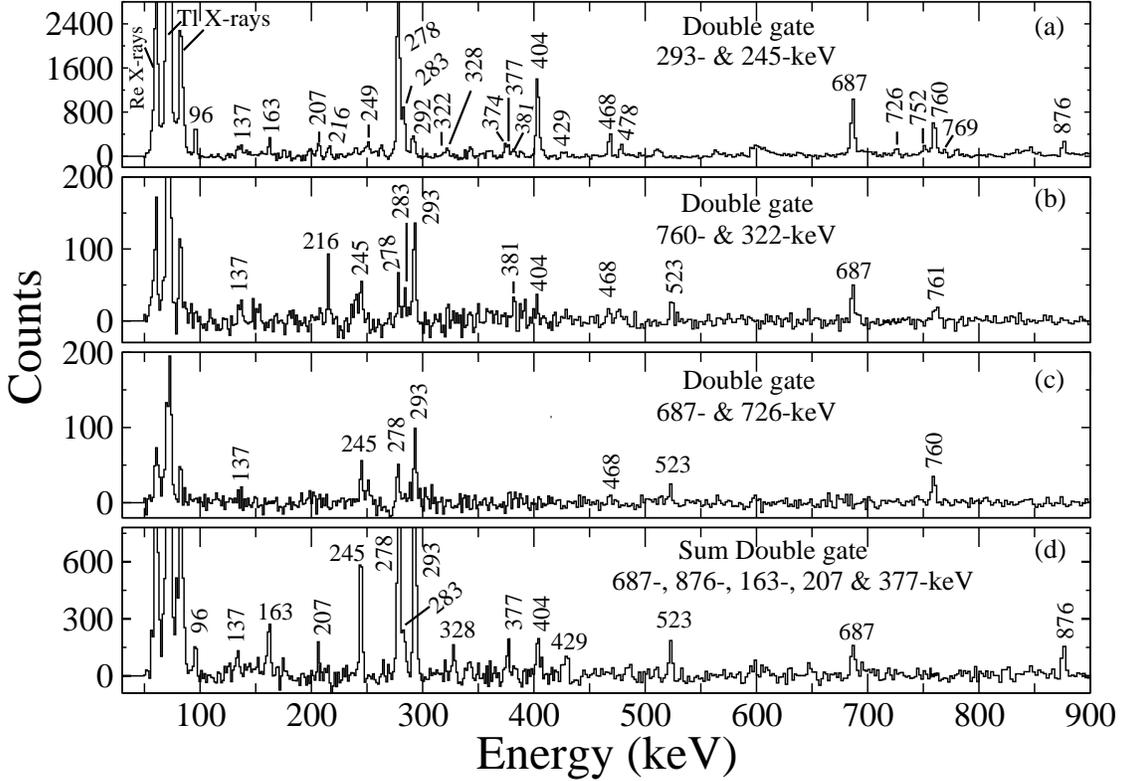}
\caption{Coincidence spectra corresponding to a double gate of (a) 293-\& 245-keV, (b) 760- \& 322-keV,
(c) 687- \& 726-keV and (d) a sum of double gated spectra corresponding to the transitions in $^{194}$Tl. }
\label{fig2}
\end{center}
\end{figure*}
%%%%%%%%%%%%%%%%%%%%%%%%%%%%%%%%%%%%%%%%%%%%%%%%%%%%%%%%%%%%

The data sorting routine ``Multi pARameter time-stamped based COincidence Search program (MARCOS)" 
developed at TIFR, sorts the time stamped data to generate a $E_{\gamma}$-$E_{\gamma}$ matrix and 
a $E_{\gamma}$-$E_{\gamma}$-$E_{\gamma}$ cube in a Radware compatible format for further analysis. 
To construct these, a coincidence time window of 400 ns was selected. The RADWARE software \cite{16} 
was used for the analysis of the matrix and the cube. The $E_{\gamma}$-$E_{\gamma}$ matrix
contained about 2.4 $\times$ 10$^9$ coincidence events. To construct the level scheme of $^{194}$Tl, 
the coincidence and the intensity relations of the $\gamma$ rays were used. Gates were put on the known 
$\gamma$-ray transitions for determining the coincidence relations. Single and double gated $\gamma$-ray spectra
from the $E_{\gamma}$-$E_{\gamma}$ matrix and $E_{\gamma}$-$E_{\gamma}$-$E_{\gamma}$ cube are shown in 
Figs. 1 and 2, respectively. The relevance of these spectra in the level scheme will be discussed in the
next section. The intensity of the $\gamma$ rays were obtained from the $E_{\gamma}$-$E_{\gamma}$ matrix 
using a single gated spectrum.

The multipolarities of the $\gamma$-ray transitions have been determined from the angular correlation 
analysis using the method of directional correlation from the oriented states (DCO) ratio, following the 
prescriptions of Kr\"{a}mer-Flecken et al.\cite{17}. For the DCO ratio analysis, the coincidence events were 
sorted into an asymmetry matrix with data from the 90$^\circ$ detectors ($\theta_1$) on one axis and -23$^\circ$ 
detectors ($\theta_2$) on the other axis. The DCO ratios (for the $\gamma$ ray $\gamma_1$, gated by a $\gamma$ ray
$\gamma_2$ of known multipolarity) are obtained from the intensities of the $\gamma$ rays (I$_\gamma$) at two 
angles $\theta_1$ and $\theta_2$, as
\begin{equation}\label{rdco}
R_{DCO} = \frac{I_{\gamma_1} ~ at~ \theta_1, ~gated ~by ~\gamma_2 ~at ~\theta_2}
               {I_{\gamma_1} ~at ~\theta_2 ~gated ~by ~\gamma_2 ~at ~\theta_1}
\end{equation}
By putting gates on the transitions with known multipolarity along the two axes of the above matrix, the DCO ratios are 
obtained for each $\gamma$ ray. For stretched transitions, the value of R$_{DCO}$ would be close to unity for the same 
multipolarity of $\gamma_1$ and $\gamma_2$. For different multipolarities and mixed transition, the value of R$_{DCO}$ 
depends on the detector angles ($\theta_1$ and $\theta_2$) and the mixing ratio ($\delta$). The validity of the R$_{DCO}$ 
measurements was checked with the known transitions in $^{194}$Tl and with the calculated values. In the present geometry, 
the calculated value of R$_{DCO}$ \cite{angcor} for a pure dipole transition gated by a stretched quadrupole transition 
is 1.65 while for a quadrupole transition gated by a pure dipole, the calculated value is 0.61. These compare well with 
the experimental values of 1.69(3) and 0.58(2), respectively, for the known pure dipole (293-keV, $E1$) and stretched 
quadrupole (687-keV, $E2$) transitions.  

%%%%%%%%%%%%%%%%%%%%%%%%%%%%%%%%%%%%%%%%%%%%%%%%%%%%%%%%%%%
\begin{figure}[ht]
\begin{center}
\includegraphics*[scale=0.3, angle = 0]{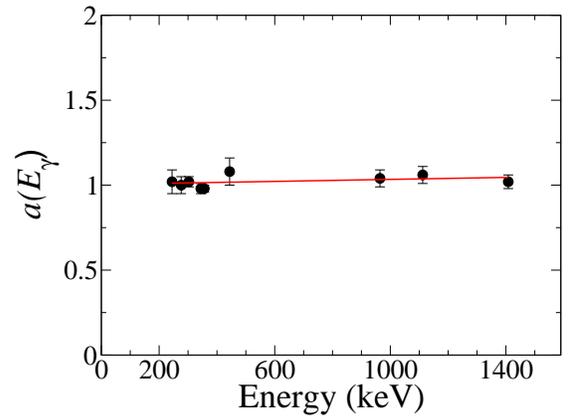}
\caption{(Color online) The asymmetry correction factor $a(E_\gamma$) at different $\gamma$ energies from $^{152}$Eu 
and $^{133}$Ba sources. The solid line corresponds to a linear fit of the data.}
\label{fig3}
\end{center}
\end{figure}
%%%%%%%%%%%%%%%%%%%%%%%%%%%%%%%%%%%%%%%%%%%%%%%%%%%%%%%%%%%%

The use of clover HPGe detectors allowed us to assign definite parities to the excited states from the measurement of 
the integrated polarization asymmetry (IPDCO) ratio, as described in \cite{18,19}, from the parallel and perpendicular 
scattering of a $\gamma$-ray photon inside the detector medium. The IPDCO ratio measurement gives a qualitative idea 
about the type of the transitions ($E/M$). The IPDCO asymmetry parameters have been deduced using the relation,
\begin{equation}\label{ipdco}
\Delta_{IPDCO} = \frac{a(E_\gamma) N_\perp - N_\parallel}{a(E_\gamma)N_\perp + N_\parallel},
\end{equation}
where $N_\parallel$ and $N_\perp$ are the counts for the actual Compton scattered $\gamma$ rays in the planes parallel 
and perpendicular to the reaction plane. The correction due to the asymmetry in the array and response of the clover 
segments, defined by $a(E_\gamma$) = $\frac{N_\parallel} {N_\perp}$, was checked using $^{152}$Eu and $^{133}$Ba sources 
and was found to be 1.004(1); which is close to unity, as expected. The data of $a(E_\gamma$) and the fitting are shown
in Fig. 3. By using the fitted parameter {$a(E_\gamma$)}, the $\Delta_{\it IPDCO}$ of the $\gamma$ rays in $^{194}$Tl 
have been determined. A positive value of $\Delta_{\it IPDCO}$ indicates an electric type transition where as a negative 
value favors a magnetic type transition. The $\Delta_{\it IPDCO}$ could not be measured for the low energy and weaker 
transitions. The low energy cut off for the polarization measurement was about 200 keV, in this work. The validity of the 
method of the IPDCO measurements was confirmed from the known transitions in $^{194}$Tl. The parallel ($N_\parallel$) and 
perpendicular ($a(E_\gamma$)*$N_\perp$) count for two $\gamma$ rays in $^{194}$Tl are shown in Fig. 4. It shows that the 
magnetic (404-keV) and the electric (687-keV) transitions can easily be identified.
%%%%%%%%%%%%%%%%%%%%%%%%%%%%%%%%%%%%%%%%%%%%%%%%%%%%%%%%%%%%
\begin{figure}[ht]
\begin{center}
\includegraphics*[scale=0.3, angle = 0]{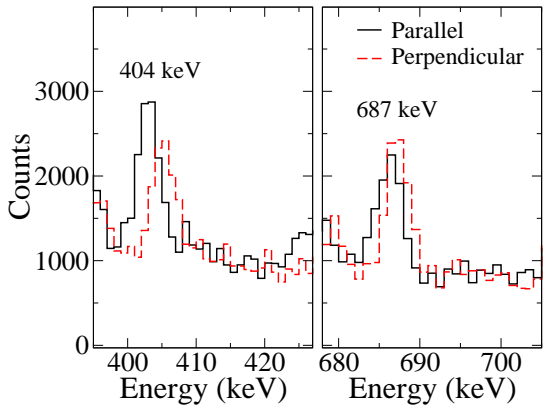}
\caption{(Color online) The perpendicular (dashed) and parallel (solid) components of the two $\gamma$ rays in $^{194}$Tl, obtained from 
the IPDCO analysis in the present work. The perpendicular component has been shifted in energy for clarity. 404-keV is a 
known magnetic type transition where as 687-keV is a known electric type transition.}
\label{fig4}
\end{center}
\end{figure}
%%%%%%%%%%%%%%%%%%%%%%%%%%%%%%%%%%%%%%%%%%%%%%%%%%%%%%%%%%%%
\begin{figure}[ht]
\begin{center}
\includegraphics*[scale=0.52, angle = -90]{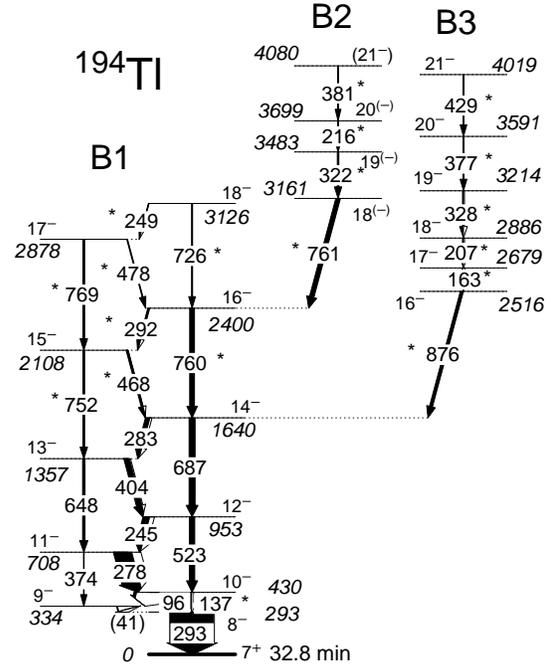}
\caption{Level scheme of $^{194}$Tl obtained from the present work. The excitation energies are given with respect to the
7$^+$ isomeric state. The new $\gamma$ rays are indicated by asterisks.}
\label{fig5}
\end{center}
\end{figure}
%%%%%%%%%%%%%%%%%%%%%%%%%%%%%%%%%%%%%%%%%%%%%%%%%%%%%%%%%%%%%

%%%%%%%%%%%%%%%%%%%%%%%%%%%%%%%%%%%%%%%%%%%%%%%%%%%%%%%%%%%%
\subsection{Experimental Results}

The level scheme of $^{194}$Tl, as obtained in the present work, is shown in Fig.~5. A total of 19 new 
$\gamma$ rays have been found and placed in the level scheme. They are marked as ``*" in the level scheme of 
Fig.~5. The deduced excitation energy, spin and parity of the excited levels and the multipolarity of the 
$\gamma$ rays, together with the other relevent information concerning their placement in the proposed level 
scheme of $^{194}$Tl, are summarized in Table I. 

\begin{table*}
\caption{\label{tab:Table1}The energy of the $\gamma$ rays ($E_{\gamma}$) observed in the present work along 
with the energy of the de-exciting levels (E$_i$), The spin and parity of the initial and final states 
($J^{\pi}_i$$\rightarrow$ $J^{\pi}_f$), intensity of the $\gamma$ rays (I$_\gamma$), the experimental values 
of the DCO ($R_{DCO}$) and the IPDCO ($\Delta_{IPDCO}$) ratios and the deduced multipolarity of the $\gamma$ rays.} 

\begin{longtable}{ccccccc}\hline

$E_{\gamma}$~~ &~~ $E_{i}$~~ &~~ $J^{\pi}_i$$\rightarrow$ $J^{\pi}_f$~~ &~~ $I_{\gamma}$
$^{\footnotemark[1]}$~~~ &~~~ $R_{DCO}$~~ &~~ $\Delta_{IPDCO}$~ &~ Deduced \\
(keV) & (keV) & & &  &  & Multipolarity \\
\hline
 
 96.1(1)  & 430.0  & $ 10^{-} $$ \rightarrow$ $9^{-} $ & 5.95(9) & 1.59(17)$^{\footnotemark[2]}$  &    -    & M1+E2 \\
 
 136.9(2) & 430.0  & $10^{-} $$ \rightarrow$ $ 8^{-}$ &2.73(6) &0.96(14)$^{\footnotemark[2]}$ &   -     &    E2 \\
 
 162.5(1)& 2679.0 & $17^{-} $$ \rightarrow$ $16^{-}$ &2.24(4)& 1.67(20)$^{\footnotemark[2]}$&   -     & M1 \\
 
 207.1(1)& 2886.1 & $18^{-} $$ \rightarrow$ $ 17^{-} $ &2.16(4)& 1.72(18)$^{\footnotemark[2]}$&   -     & M1    \\
 
 215.5(2) & 3698.6 & $20^{(-)}  $$ \rightarrow$ $19^{(-)}$ &1.40(2) & 1.51(14)$^{\footnotemark[2]}$& -     &M1+E2 \\
 
 244.9(1) &  953.2 & $12^{-}  $$ \rightarrow$ $11^{-} $ &14.92(41)& 1.61(5)$^{\footnotemark[2]}$&-0.10(2) & M1(+E2) \\
 
 248.6(3) & 3125.5 & $18^{-} $$ \rightarrow$ $17^{-}$ &0.94(2) & 1.64(19)$^{\footnotemark[2]}$&-0.13(10)& M1+E2\\

 278.4(1) &  708.4& $11^{-}  $$ \rightarrow$ $10^{-}$ &41.86(62)& 1.62(3)$^{\footnotemark[2]}$  &-0.14(1) & M1(+E2)  \\

 283.2(1) & 1640.0 & $14^{-} $$ \rightarrow$ $13^{-}$ &10.00(15)&1.85(14)$^{\footnotemark[3]}$ &-0.24(4)& M1(+E2)\\

 291.9(2) & 2399.6 & $16^{-} $$ \rightarrow$ $15^{-}$ &3.06(7) &1.46(10)$^{\footnotemark[4]}$ &-& M1+E2 \\

 293.1(1) & 293.1  & $8^{-} $$ \rightarrow$ $7^{+}$ &100.0(4)& 1.69(3)$^{\footnotemark[2]} $& 0.07(2)& E1\\

 322.2(1)& 3483.0 & $19^{(-)} $$ \rightarrow$ $18^{(-)}$ &2.04(6)& 1.13(9)$^{\footnotemark[4]}$ &-        & M1+E2 \\

 327.7(2)& 3213.8 & $19^{-} $$ \rightarrow$ $18^{-}$ &3.37(5)& 1.71(23)$^{\footnotemark[2]}$ &-0.25(6)    & M1 \\

 373.8(2) & 708.4  & $11^{-}$$ \rightarrow$ $9^{-}$ &1.17(4)&0.95(14)$^{\footnotemark[2]}$ &  -    &E2\\

 376.9(1) &  3590.7 & $20^{-}$$ \rightarrow$ $19^{-}$ &1.91(5)& 1.67(21)$^{\footnotemark[2]}$   & -0.22(8)  &  M1 \\

 381.3(2) &  4079.8 & $(21^{-}$)$ \rightarrow$ $20^{(-)}$ &0.40(4)& -   &   -     &               -     \\

 403.5(1)& 1356.7 & $13^{-}$$ \rightarrow$ $12^{-}$ &12.67(19)& 1.85(5)$^{\footnotemark[4]}$  &-0.05(2)&M1+E2\\

 428.6(2)& 4019.3 & $21^{-} $$ \rightarrow$ $20^{-}$ &1.03(2)& 1.69(37)$^{\footnotemark[2]}$ &-        & M1 \\

 468.4(1) &2108.5 & $15^{-} $$ \rightarrow$ $14^{-}$ &4.13(7)& 1.50(7)$^{\footnotemark[2]}$   &-0.20(8) & M1+E2 \\
 
 478.3(2) & 2877.6 & $17^{-} $$ \rightarrow$ $16^{-}$ &1.70(3)& 1.73(24)$^{\footnotemark[2]}$   &-0.13(10) & M1+E2\\

 523.1(1) &953.2   & $12^{-} $$ \rightarrow$ $10^{-} $ & 11.87(18)&0.68(2)$^{\footnotemark[4]}$ &0.08(3)  & E2 \\

 648.3(1) &1356.7  & $13^{-} $$ \rightarrow$ $11^{-} $ & 5.42(8)&0.65(4)$^{\footnotemark[4]}$ &0.39(5)  & E2 \\

 686.7(1) &1640.0  & $14^{-} $$ \rightarrow$ $12^{-} $ & 13.62(20)&0.58(2)$^{\footnotemark[4]}$ &0.08(4)  & E2 \\

 725.8(3) &3125.5  & $18^{-} $$ \rightarrow$ $16^{-} $ & 2.27(6)&0.57(6)$^{\footnotemark[4]}$ &      -  & E2 \\

 751.8(2) &2108.5  & $15^{-} $$ \rightarrow$ $13^{-} $ & 3.80(6)&0.54(4)$^{\footnotemark[4]}$ &0.27(8)  & E2 \\

 759.5(2) &2399.6  & $16^{-} $$ \rightarrow$ $14^{-} $ & 10.84(24)&1.03(14)$^{\footnotemark[5]}$ &0.30(10)  & E2 \\

 761.1(2) &3160.8  & $18^{(-)} $$ \rightarrow$ $16^{-} $ & 10.23(43)&0.58(9)$^{\footnotemark[6]}$ &-  & E2 \\

 769.1(2) &2877.6  & $17^{-} $$ \rightarrow$ $15^{-} $ & 3.87(6)&0.68(3)$^{\footnotemark[4]}$ &0.07(4)  & E2 \\

 876.4(1) &2516.4  & $16^{-} $$ \rightarrow$ $14^{-} $ & 6.90(11)&0.67(4)$^{\footnotemark[4]}$ &0.11(3)  & E2 \\

\hline
\end{longtable}
\footnotetext[1]{Relative $\gamma$-ray intensities normalized to 100 for the 293.1-keV $\gamma$ ray.}
\footnotetext[2]{From 686.7-keV ($E2$) DCO gate.}
\footnotetext[3]{From 648.3-keV ($E2$) DCO gate.}
\footnotetext[4]{From 293.1-keV  ($E1$) DCO gate.}
\footnotetext[5]{From 725.8-keV ($E2$) DCO gate.}
\footnotetext[6]{From 468.4-keV ($M1$) DCO gate.}
\end{table*}

%%%%%%%%%%%%%%%%%%%%%%%%%%%%%%%%%%%%%%%%%%%%%%%%%%%%%%%%%%%%%%%%%%%%%%%%%%%%%%%%%%%%%%%%%%%%%%%%%%

The 2$^-$ ground state (not shown in Fig. 5) and the 7$^+$ isomeric state, (T$_{1/2} = $32.8 min) were 
known in $^{194}$Tl from the beta decay studies \cite{ensdf}. No $\gamma$-ray transition was known to decay 
from the isomeric state. The excitation energy of this state was also not known. Both the ground and the 
isomeric states decay by electron capture decay. The excitation energies of the states, in the level scheme 
presented in Fig.~5, have been given with respect to the 7$^+$ isomeric state as was presented in Refs. \cite{7,13} 
for $^{190,198}$Tl. In the present work, the level scheme of $^{194}$Tl has been extended up to an excitation 
energy of $\sim 4.1$ MeV and (21$^-$)$\hbar$ spin and is a much improved one compared to the previouly known 
level scheme reported in Ref. \cite{9}. The proposed level scheme is based on the following arguments.

A 293-keV $\gamma$ ray from the 8$^-$ to the 7$^+$ isomeric state in $^{194}$Tl was known, from the systematics 
of the Tl isotopes, to be a hindered E1 transition. The DCO ratio and the positive value of the IPDCO ratio, 
obtained in the present work, confirm this assignment. A 137-keV $\gamma$ ray has been observed in the present 
work which is in coincidence with the known 293-keV transition from the 8$^-$ to the 7$^+$ state as can be seen 
from Fig. 1 (top panel). A spectrum gated by this 137-keV $\gamma$ line, shown in the bottom panel of Fig. 1, shows 
all the $\gamma$ rays in $^{194}$Tl present in the 293-keV gated spectrum, except for the lines at 96 keV and 374 keV. 
This $\gamma$ ray is also observed in the double gated spectra shown in Fig. 2. This clearly indicates that there is a 
137 keV $\gamma$-ray transition above the 293-keV transition and in parallel to the 96-keV transition. It established the 
334-keV level as the previously unknown energy upon which the rest of the band is built. The DCO ratio for the 
137-keV $\gamma$ ray has been measured in this work gated by the 687-keV $\gamma$ ray which was known to be a quadrupole 
($E2$) transition \cite{9}. The value of this ratio comes out to be close to unity (see Table-I) indicating the stretched 
quadrupole nature of the 137-keV transition. Although, the energy of this $\gamma$ ray is too low for the IPDCO measurement 
but the M2 assignment for this low-energy transition may be completely ruled out from the lifetime consideration. Therefore, 
considering the 137-keV $\gamma$ ray as an $E2$ transition, spin and parity of the 430-keV level has been assigned, in
this work, as 10$^-$. 
The 374-keV $\gamma$ ray was reported as a tentative one in Ref.\cite{9}. The double gated spectrum of 
293- and 245-keV shown in Fig. 2(a) has a peak at 374 keV which confirms its placement. The $E2$ nature of this
transition is also evident from the measured R$_{DCO}$ value (close to unity) gated by a known $E2$ transition.
The placement of the above $\gamma$ rays indicates that there should be a 41-keV transition from the 9$^-$ to the 
8$^-$ state. This low energy highly converted ($\alpha \sim 200$) transition has not been observed in this work 
as our experimental set up was not suited to detect such transitions.

It can be seen from Fig. 2(a) that all the $\gamma$ rays reported in Ref. \cite{9} have been observed in the present 
work up to the 1640-keV state. However, the 748.6-, 741.9-, 458.6-, 289.4- keV $\gamma$ rays, reported in Ref. \cite{9} and 
placed above the 1640-keV state, were not observed. Instead, we have observed a cascade of 468-, 292-, 478- and 
249-keV $M1+E2$ $\gamma$ rays above the 1640-keV state along with the cross-over $E2$ transitions. These $\gamma$ rays have 
been placed above the 687-keV $\gamma$ ray in band B1 as they satisfy the corresponding coincidence relations. 
It may be pointed out that a more efficient experimental set up was used in the present work than the earlier work. 
With the observation of these transitions, the negative parity ground band B1 has been extended up to an excitation 
energy of 3.13 MeV and a spin of 18$\hbar$.

We have also observed two other band like structures B2 and B3, for the first time in $^{194}$Tl. They are placed 
above the 16$^-$ and the 14$^-$ states of the band B1, respectively. The $\gamma$ lines belonging to these bands have
been seen in the single and double gated spectra shown in Fig. 1 and Fig. 2(a). The spectrum in Fig. 2(b) (double 
gate on 760- and 322-keV)
shows the $\gamma$ rays belonging to the band B2. The spectrum also shows the 761-keV $\gamma$ ray, indicating its double 
placement. The 687-keV line is observed in this spectrum but not the 726-keV line. The 216- and the 381-keV $\gamma$ rays, 
belonging to this band, are also observed in this spectrum. The double gated spectrum of 687- and 726-keV [Fig. 2(c)] 
shows the 760-keV line, slightly lower in energy than the one in Fig. 2(b). Therefore, the higher energy transition 
(i.e 761-keV) of the 760-keV doublet has been assigned as the linking transition between the bands B2 and B1.  
Therefore, the lower energy transition (i.e 760-keV) of the doublet corresponds to the member of the band B1. The 
216-, 322- and 381-keV $\gamma$ rays are not observed in the spectrum of Fig. 2(c) and hence, they form the members of 
the band B2. 

To get an unambiguous value for the DCO ratios for the 760-and 761-keV transitions, the gating transitions
were selected carefully. The DCO ratio values of the doublet cannot be distinguised for any choice of a pure 
$E2$ gating transition below the 14$^-$ level. However, in the 468-keV gate, the 760-keV transition (in the band B1) 
would be absent and an unambiguous value of DCO ratio can be obtained for the 761-keV linking transition. The
468-keV transition has been found to be an $M1$ transition with very little or no $E2$ admixture from the negative
value of its IPDCO ratio. So, the 468-keV dipole transition was chosen to obtain the DCO ratio of the 761-keV linking 
transition and similarly, the 726-keV quadrupole transition has been chosen as the gating transition to obtain the DCO 
ratio value of the 760-keV in-band transition. The DCO ratios of the 760- and 761-keV transitions, obtained in this way, 
suggest that both of them are quadrupole transitions. The IPDCO ratio, measured for the in-band 760-keV transition, clearly 
suggests that it is an $E2$ transition. The IPDCO ratio could not be obtained for the 761-keV transition and an 
$E2$ character for this linking transition has been assumed to assign the parity of the band B2, tentatively, as 
negative. The $M2$ character for the 761-keV transition could not, however, be ruled out and, hence the parity of 
the band B2 could be positive as well. A lifetime measurement of the 18$^{(-)}$ state is necessary to overcome this 
ambiguity. The Weisskopf estimate of the halflives corresponding to $E2$ and $M2$ character of this transition
would be $\sim 50$ pico-sec and $\sim 5$ nsec, respectively. In the present work, we could not distinguish these two
halflives.

The 876-keV $\gamma$ ray has been observed in the double gated spectrum in Fig. 2(a) but not in the double gated 
spectrum in Fig. 2(b) or 2(c). This $\gamma$ ray was observed with any combination of double gates below the 
1640-keV level but was not observed with any $\gamma$ ray above this level. This indicates that the 876-keV $\gamma$ 
ray must decay to the 14$^-$ state of the band B1. The cascade of $\gamma$ rays in the band B3 has been observed 
in Fig. 2(a). A sum of double gated spectra, constructed from the combination of the 687-, 876-, 163-, 207 \& 
377-keV $\gamma$ rays, is shown in Fig. 2(d).  This spectrum clearly shows the 163-, 207-, 328-, 377- \& 429-keV 
$\gamma$ rays belonging to this band. The $E2$ nature of the 876-keV $\gamma$ ray is evident from its DCO and IPDCO 
ratios. Therefore, the bandhead of the band B3 has been assigned as 16$^-$. The ordering of the $\gamma$s in this 
band are based on their total intensities. However, since the cross over $E2$ transitions in this band were
not observed, the ordering of the levels may be considered as somewhat tentative.
The spins and parities of the states in this band are assigned from the measured DCO and the IPDCO ratios of the 
$\gamma$-ray transitions, wherever possible. The R$_{DCO}$ values of these transitions are very close to the expected 
values for pure dipole transitions. The negative values obtained for the IPDCO ratios for the 328- and the 377-keV 
transitions, together with their R$_{DCO}$ values, give clear evidence that they are predominantly M1 in nature. 
Since the other $\gamma$ rays are in-band with the 328- and the 377-keV M1 transitions, so, we have assigned 
M1 nature for all the $\gamma$ rays in this band.

\section{Discussion}

The high spin band structure in $^{194}$Tl, obtained in the present work, can be discussed in the light of the
neighboring isotopes. Ground state band structures in the odd-odd Tl isotopes in $A\sim 190$ region are similar
to band B1 in $^{194}$Tl. $^{190}$Tl and $^{198}$Tl are the other two isotopes for which definite excitation 
energies, spins and parities are known for this band. However, it is interesting that the structure of this band 
in these two nuclei has been interpreted differently. In $^{190}$Tl, this band was interpreted with oblate 
deformation \cite{7} while in $^{198}$Tl, a possible chiral structure has been reported with triaxial deformation 
\cite{13}. We have compared different properties of the ground state band in the odd-odd Tl isotopes and have shown 
that they look very similar. 

The configurations of the first few excited states in $^{194}$Tl are attributed by the following arguments.
The ground state of the odd-A Tl isotopes are 1/2$^+$ corresponding to the 3$s_{1/2}$ orbital near the proton 
Fermi level. In the isotone $^{193}$Hg, the experimentally observed 5/2$^-$ and 13/2$^+$ states lie very close 
(within 40- and 141-keV, respectively) to the 3/2$^-$ ground state \cite{ensdf1}. These states correspond to the 
$f_{5/2}$, $i_{13/2}$ and $p_{3/2}$ orbitals, respectively, near the neutron Fermi level. Correspondingly, the 
configurations of the ground state (2$^-$) and the 7$^+$ isomeric state in the odd-odd $^{194}$Tl nucleus have 
been attributed to $\pi s_{1/2} \otimes \nu p_{3/2}$ and $\pi s_{1/2} \otimes \nu i_{13/2}$ configurations, 
respectively. The 8$^-$ state, at the excitation energy of 293-keV, in this nucleus, has been interpreted as the 
band head of the $\pi h_{9/2} \otimes \nu i_{13/2}$ configuration \cite{9}. These orbitals lie near the proton 
and the neutron Fermi levels of $^{194}$Tl for an oblate deformation. The excitation energies of the 9$^-$, 
10$^-$ and 11$^-$ states in this band for $^{190,194,198}$Tl nuclei are shown in Fig. 6, with respect to the 
excitation energy of the 7$^+$ isomeric state. It can be seen that the energies of these states increase smoothly 
with mass number A, as expected.

%%%%%%%%%%%%%%%%%%%%%%%%%%%%%%%%%%%%%%%%%%%%%%%%%%%%%%%%%%%
\begin{figure}[ht]
\begin{center}
\includegraphics*[scale=0.25, angle = 0]{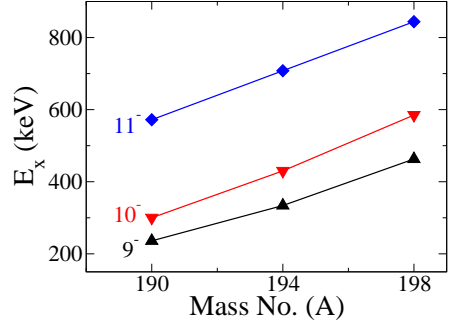}
\caption{(Color online) Excitation energies of the 9$^-$, 10$^-$ and 11$^-$ states in odd-odd isotopes of Tl as a
function of mass No. A. Only the data for those isotopes are plotted for which the definite excitation
energies are known with respect to the 7$^+$ isomeric state. }
\label{fig6}
\end{center}
\end{figure}
%%%%%%%%%%%%%%%%%%%%%%%%%%%%%%%%%%%%%%%%%%%%%%%%%%%%%%%%%%%%

The alignment ($i_x$) for the band B1 in $^{194}$Tl is shown in Fig.~7 as a function of the rotational frequency 
$\hbar\omega$. The alignments for the neighboring odd-A nuclei $^{193}$Tl and $^{193}$Hg are also shown in Fig.~7 which
have been deduced from the corresponding level schemes reported by Reviol et al. \cite{22} and Hubel et al. \cite{6}, 
respectively. It can be seen that the band in $^{194}$Tl has an initial alignment of about 5$\hbar$. This value is in
good agreement with the value obtained from the initial alignments of the neighboring odd-A nuclei using the additivity 
rule \cite{20,21}. There is a gain in alignment of about 5$\hbar$ at rotational frequency of $\hbar\omega\sim 0.34$ MeV
which is similar to $^{190}$Tl and $^{198}$Tl. The similarity between these odd-odd Tl isotopes are also reflected in 
their kinetic moments of inertia (J$^{(1)}$) values as shown in Fig. 8. This indicates similar deformation 
in these nuclei. The particle alignments in these isotopes are also taking place at about the same frequency. In the 
present work, the band B1 in $^{194}$Tl could be extended just beyond the band crossing at $\hbar\omega_c \sim 0.34$ MeV.

%%%%%%%%%%%%%%%%%%%%%%%%%%%%%%%%%%%%%%%%%%%%%%%%%%%%%%%%%%%
\begin{figure}[ht]
\begin{center}
\includegraphics*[scale=0.3, angle = 0]{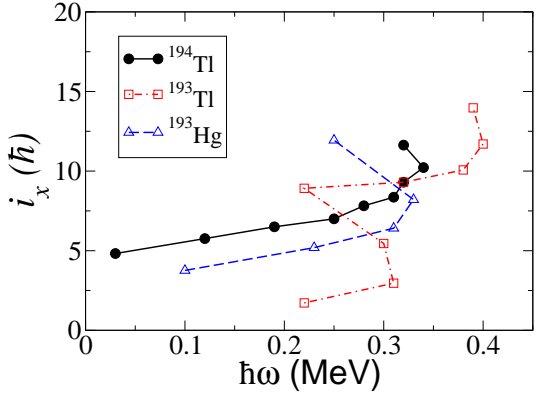}
\caption{(Color online) Experimental alignments ($i_x$) as a function of the rotational frequency ($\hbar\omega$) for 
the $\pi h_{9/2}$ band in $^{193}$Tl \cite{22}, $\nu i_{13/2}$ band in $^{193}$Hg \cite{6} and band B1 in $^{194}$Tl. 
The Harris reference parameters are chosen to be $J_{0}$ = 8.0{$\hbar$}$^{2}$ $MeV^{-1}$ and $J_{1}$ = 40{$\hbar$}$^{4}$ 
$MeV^{-3}$.}
\label{fig7}
\end{center}
\end{figure}
%%%%%%%%%%%%%%%%%%%%%%%%%%%%%%%%%%%%%%%%%%%%%%%%%%%%%%%%%%%

The band crossing phenomena in the odd-odd Tl isotopes can be understood from the band crossings in the even-even 
core of Hg isotopes and the band crossings in the odd-A Tl isotopes. In $^{192}$Hg \cite{6}, the first two observed 
band crossings, at $\hbar\omega_{c1} \sim 0.210$ MeV and at $\hbar\omega_{c2} \sim 0.362$ MeV, were interpreted as 
due to the alignments of the $i_{13/2}$ neutrons (G $\rightarrow$ AB and AB $\rightarrow$ ABCD crossing, following 
the nomenclature of Hubel {\it et al.} \cite{6}). The alignment of the protons takes place at even higher frequencies. 
The band crossings in the odd-Z $^{193}$Tl nucleus, in which the first proton crossing is blocked, were also interpreted 
as due to the alignments of the same $i_{13/2}$ neutrons. The crossing frequency of $\hbar\omega_c \sim 0.34$ MeV in 
$^{194}$Tl agrees well with the $\hbar\omega_{c2}$ value in $^{192}$Hg. Therefore, the observed band crossing in 
doubly-odd $^{194}$Tl may be attributed to the alignment of a pair of neutrons in the $i_{13/2}$ orbital. The similarity 
and the systematic trend of the observed AB $\rightarrow$ ABCD crossing frequencies for $^{190}$Hg ($\hbar\omega_{c2} 
\sim 0.352$ MeV), $^{192}$Hg ($\hbar\omega_{c2} \sim 0.362$ MeV) and $^{194}$Hg ($\hbar\omega_{c2} \sim 0.348$ MeV) 
\cite{6} are also in good agreement with the crossing frequencies of $^{192}$Tl ($\hbar\omega_c \sim 0.32$ MeV), 
$^{194}$Tl ($\hbar\omega_c \sim 0.34$ MeV) and $^{196}$Tl ($\hbar\omega_c \sim 0.31$ MeV).

%%%%%%%%%%%%%%%%%%%%%%%%%%%%%%%%%%%%%%%%%%%%%%%%%%%%%%%%%%%
\begin{figure}[ht]
\begin{center}
\includegraphics*[scale=0.3, angle = 0]{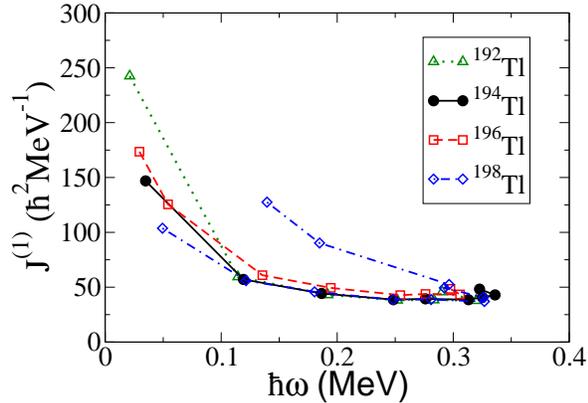}
\caption{(Color online) Experimental values of the moments of inertia J$^{(1)}$ as a function of the rotational frequency ($\hbar\omega$) 
for the band B1 in $^{194}$Tl are compared with the similar bands in $^{192,196,198}$Tl.} 
\label{fig9}
\end{center}
\end{figure}
%%%%%%%%%%%%%%%%%%%%%%%%%%%%%%%%%%%%%%%%%%%%%%%%%%%%%%%%%%%

%%%%%%%%%%%%%%%%%%%%%%%%%%%%%%%%%%%%%%%%%%%%%%%%%%%%%%%%%%%
\begin{figure}[ht]
\begin{center}
\includegraphics*[scale=0.3, angle = 0]{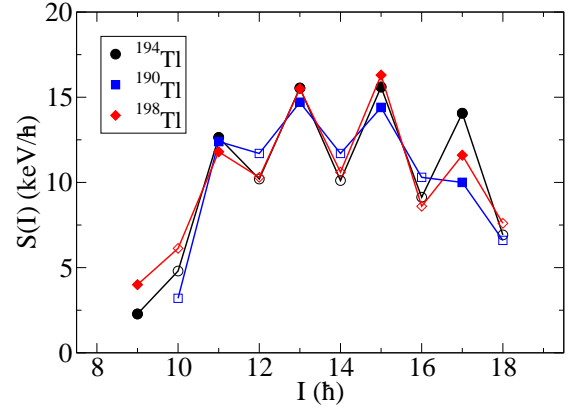}
\caption{(Color online) The staggering, S(I) = [E(I) - E(I-1)]/2I, plots as a function of spin (I) for the negative parity 
yrast bands in $^{194}$Tl along with those in $^{190,198}$Tl.}
\label{fig10}
\end{center}
\end{figure}
%%%%%%%%%%%%%%%%%%%%%%%%%%%%%%%%%%%%%%%%%%%%%%%%%%%%%%%%%%%

A low spin signature inversion has been observed in the rotational bands in odd-odd nuclei in various mass
regions, involving high-j configurations. In particular, the signature inversion has been reported for the 
prolate deformed odd-odd nuclei in the rare earth region with a $\pi h_{9/2} \otimes \nu i_{13/2}$ configuration 
\cite{stag1}. In the A $\sim$ 190 region, the low-spin signature inversion for a band associated with an oblate 
$\pi h_{9/2} \otimes \nu i_{13/2}$ configuration has been reported for the first time in $^{190}$Tl \cite{7}. In 
the present work, a similar signature inversion has also been observed in $^{194}$Tl. The 
signature inversion in a band can be identified experimentally in the plot of energy staggering, defined by 
$S(I) = [E(I) - E(I-1)]/2I$, where $E(I)$ is the energy of the state with spin $I$, as a function of the spin. 
The same has been plotted in Fig. 9 for the negative parity yrast band of $^{194}$Tl along with those for 
$^{190,198}$Tl. The staggering plots of these three nuclei show remarkably similar behavior with low-spin 
signature inversion at the same spin value of 11$\hbar$. A J-dependent residual $p-n$ interaction was used to 
interpret the signature inversion in $^{190}$Tl \cite{7}. The similarity in the staggering plots suggest that the 
residual $p-n$ interaction remains almost unchanged for the heavier odd-odd isotopes of $^{194,198}$Tl.

The similarities in the band crossing frequencies, the moments of inertia, the signature inversion and the 
staggering plots for the above three nuclei suggest that the high spin properties of the ground state band 
in all the odd-odd Tl isotopes in $A\sim 190$ region are very similar and therefore, they are expected to have 
similar structure. In the present work, we have not observed any indication of a chiral side band in $^{194}$Tl 
unlike in $^{198}$Tl.

The band B3 in $^{194}$Tl is a negative parity band with band head spin of 16$\hbar$ at an excitation energy of 
2.5 MeV. This band has been extended up to 21$\hbar$ with predominnatly $M1$ transitions. No $E2$ cross over transitions 
could be observed in this band. The band B3 lies at an energy which is somewhat higher than the crossing of the 
2- and 4-quasiparticle configurations of the band B1. Therefore, the band B3 seems to be a 6-quasiparticle band with 
two more protons in the $\pi$h$_{9/2}$ and $\pi$s$_{1/2}$ orbitals. The configuration of the 
band B3 could, therefore, be assigned as $\pi h_{9/2}^2 s_{1/2}^{-1} \otimes \nu i_{13/2}^{-2} p_{3/2}$. The respective 
proton and neutron orbitals constitute the 2-quasiparticle states in this nucleus at the lower excitation energies (2$^-$ 
state, 7$^+$ state and the 8$^-$ band head of the band B1). The experimental initial aligned angular momentum of 7$\hbar$, 
deduced for this band, is consistent with the above configuration. The particle-hole configuration of this 
band with proton particles in high-j, high-$\Omega$ and neutron holes in high-j, low-$\Omega$ orbitals (for an oblate 
deformation) is favorable for magnetic rotation (MR) and therefore, the excited states, in this band, may be generated by 
the shear mechanism. This band, with 3 particle-hole pairs, seems to follow the same general features of MR bands 
\cite{15,mrband}. For the MR bands, the level energies ($E$) and the spin ($I$) in the band follow the pattern 
$E - E_0 \sim A(I - I_0)^{2}$ where, E$_0$ and I$_0$ are the energy and the spin of the band head, respectively. 
The plot of $E - E_0$ vs. $(I - I_0)^2$ for this band is shown in Fig. 10. The solid line is the fit of the data 
using the above relation. The good agreement of the data with the fitted curve, in this plot, clearly indicates 
that the band B3 closely follows the above relation. Considering a higher limit of the intensity of the unobserved 
crossover $E2$ transitions as the level of the background in our data, the lower limit of the $B(M1)/B(E2)$ ratio 
has been estimated to be $> $28 $\mu^2/(eb)^2$. This compares well with the typical value of $\ge $20 $\mu^2/(eb)^2$ 
for a MR band. The dynamic moment of inertia J$^{(2)} \sim 24$ $\hbar^2$ MeV$^{-1}$ obtained for this band is also 
within the typical value of J$^{(2)} \sim 10 - 25$ $\hbar^2$ MeV$^{-1}$ for an MR band. All these indicate that the 
band B3 is, most likely, an MR band.  

%%%%%%%%%%%%%%%%%%%%%%%%%%%%%%%%%%%%%%%%%%%%%%%%%%%%%%%%%%%
\begin{figure}[ht]
\begin{center}
\includegraphics*[scale=0.28, angle = 0]{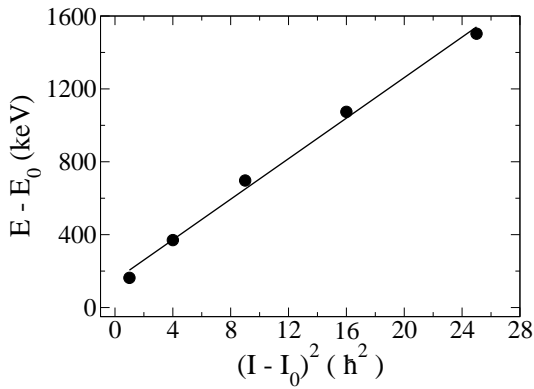}
\caption{Relative energy (E) vs. spin (I) curve for the band B3 built on the $16^{-}$ band-head. E$_o$ and I$_o$ are the 
band head energy and spin respectively. The fitted curve is shown by the solid lines (see text for details).} 
\label{fig10}
\end{center}
\end{figure}
%%%%%%%%%%%%%%%%%%%%%%%%%%%%%%%%%%%%%%%%%%%%%%%%%%%%%%%%%%%

The band like structure B2 lies at an excitation energy of 3.16 MeV with the band head spin of 18$\hbar$. The
parity of this band remains tentative and this band is also not well developed. Therefore, characterization of this 
band at this moment would be rather prematured. However, from the systematics of the odd-A Tl nuclei, it appears that 
this band may be generated with an additional proton pair in the i$_{13/2}$ orbital with the 4-quasiparticle configuration 
of the band B1 above its band crossing.

\subsection{TRS calculations}

In order to get an idea about the shape of the Tl nuclei for different configurations, the total Routhian 
surface (TRS) calculations have been performed, in the $\beta_2 - \gamma$ deformation mesh points, where
$\beta_2$ and $\gamma$ are the quadrupole deformation parameters, with 
minimization in the hexadcapole deformation $\beta_4$, for the bands B1 and B3 in $^{194}$Tl. The 
Hartee-Fock-Bogoliubov code of Nazarewicz {\it et al.} \cite{naza1,naza2} was used for the calculations. 
A deformed Woods-Saxon potential and pairing interaction were used with the Strutinsky shell corrections 
method. The procedure has been outlined in Refs. \cite{gm1,gm2}. The contour plots of the TRSs are shown 
in Fig. 11 for the 2-quasiparticle $\pi h_{9/2} \otimes \nu i_{13/2}$ configuration corresponding to the 
band B1 (top panel) and for the 6-quasiparticle $\pi h_{9/2}^2 s_{1/2}^{-1} \otimes \nu i_{13/2}^{-2} p_{3/2}$ 
configuration corresponding to band B3 (bottom panel) in $^{194}$Tl. These surfaces were calculated at the 
rotational frequencies of $\hbar\omega = 0.11$ MeV and 0.16 MeV, respectively. The plots clearly show  
minima in the TRS at an oblate deformation ($\gamma = 0^o$ is prolate and $\gamma = -60^o$ is oblate) with 
$\beta_2 \sim 0.15$ and $\gamma \sim -57^o$ for the band B1 and a near spherical shape with $\beta_2 \sim 0.06$ 
and $\gamma \sim -80^o$ for the band B3. 

The surfaces calculated for the same 2-quasiparticle configuration as in band B1 are shown in Fig. 12 for 
$^{196}$Tl and in $^{198}$Tl. The minima for these nuclei are also found to be at the oblate deformation 
with very similar $\beta_2$ values. The similarity in the calculated shapes for these nuclei corroborates 
well with the observed similarities in the properties of the $\pi h_{9/2} \otimes \nu i _{13/2}$ bands in
odd-odd Tl nuclei. The oblate shapes in all these three nuclei are predicted to persist over a rotational 
frequency range up to the band crossing. After the band crossing, the shape of $^{194}$Tl remains almost 
the same with deformation $\beta_2 \sim 0.13$ and $\gamma \sim -53^o$. The lack of quadrupole deformation 
calculated for the band B3 in $^{194}$Tl is consistent with the non-observation of $E2$ transitions and the 
conjecture of MR nature of this band. 

%%%%%%%%%%%%%%%%%%%%%%%%%%%%%%%%%%%%%%%%%%%%%%%%%%%%%%%%%%%
\begin{figure}[ht]
\begin{center}
\includegraphics*[scale=0.55,  angle =0]{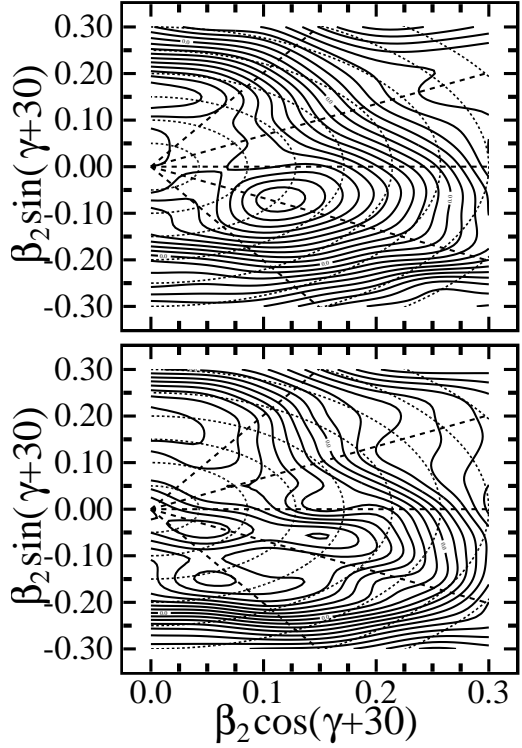}
\caption{Contour plots of the total Routhian surfaces (TRSs) in the $\beta_2$ - $\gamma$ deformation mesh
for the configurations of the bands B1 (top) and B3 (bottom) in $^{194}$Tl at the rotational frequencies
$\hbar\omega = 0.11$ MeV and 0.16 MeV, respectively. The contours are 400 keV apart}
\label{fig11}
\end{center}
\end{figure}
%%%%%%%%%%%%%%%%%%%%%%%%%%%%%%%%%%%%%%%%%%%%%%%%%%%%%%%%%%%

%%%%%%%%%%%%%%%%%%%%%%%%%%%%%%%%%%%%%%%%%%%%%%%%%%%%%%%%%%%
\begin{figure}[ht]
\begin{center}
\includegraphics*[scale=0.55,  angle =0]{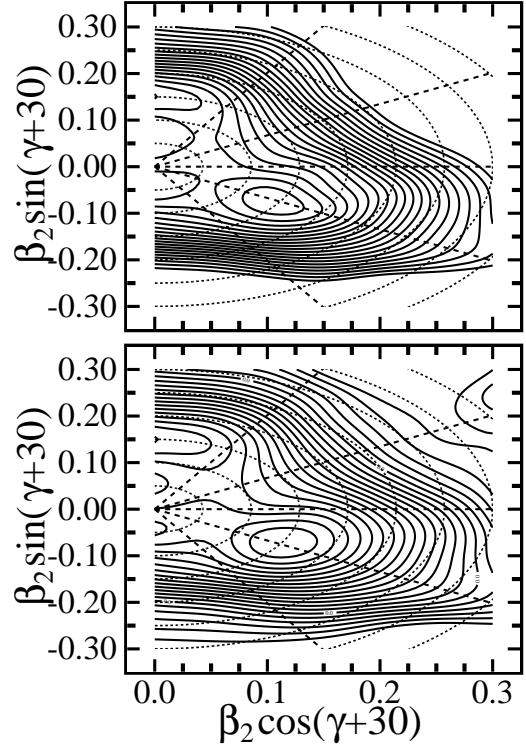}
\caption{Same as Fig. 11 but for $^{196}$Tl (bottom) and $^{198}$Tl (top) for the same configuration
as band B1 of $^{194}$Tl.}
\label{fig12}
\end{center}
\end{figure}
%%%%%%%%%%%%%%%%%%%%%%%%%%%%%%%%%%%%%%%%%%%%%%%%%%%%%%%%%%%

%%%%%%%%%%%%%%%%%%%%%%%%%%%%%%%%%%%%%%%%%%%%%%%%%%%%%%%%%%%
\begin{figure}[ht]
\begin{center}
\includegraphics*[scale=0.32,  angle =0]{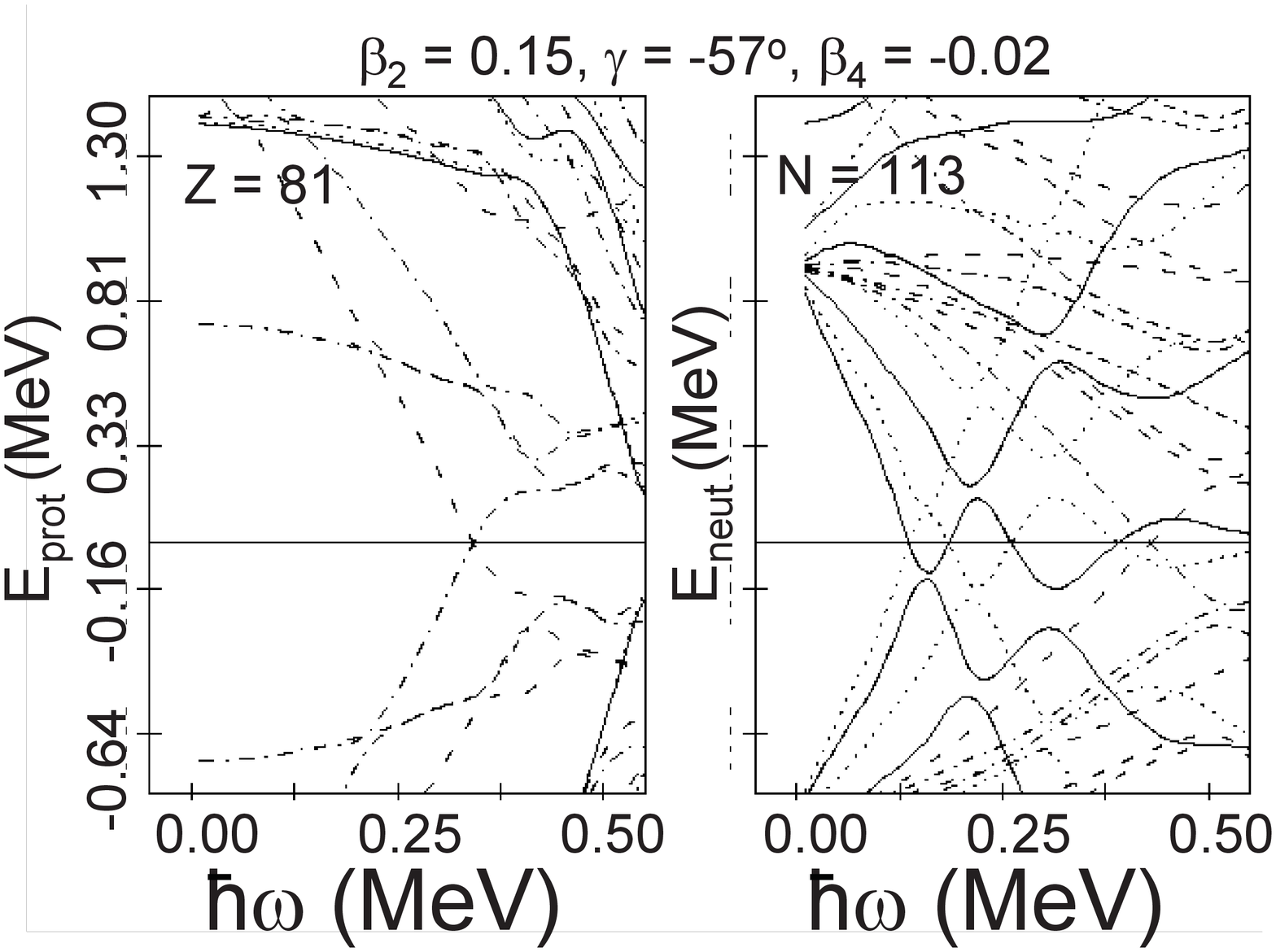}
\caption{Calculated quasiparticle Routhians as a function of rotational frequency $\hbar\omega$ for
Z = 81 (left) and N = 113 (right) for the deformation $\beta_2 = 0.15$, $\gamma = -57^o$ and
$\beta_4 = -0.02$. The quantum numbers ($\pi, \alpha$) of the levels are drawn as: solid line
(+,+1/2), dotted line (+,-1/2), dash-dotted line(-,+1/2) and dashed line (-,-1/2).}
\label{fig13}
\end{center}
\end{figure}
%%%%%%%%%%%%%%%%%%%%%%%%%%%%%%%%%%%%%%%%%%%%%%%%%%%%%%%%%%%

The calculated quasiparticle Routhians for the protons (Z = 81) and the neutrons (N = 113) in $^{194}$Tl are 
shown in left and right panels, respectively, of Fig. 13 for the deformation parameters $\beta_2 = 0.15,
\gamma = -57^o$ and $\beta_4 = -0.02$ corresponding to the minimum in the TRS in Fig. 11. These calculations 
predict the first neutron pair alignment (AB crossing) at around $\hbar\omega \sim 0.14$ MeV and a second one
(CD crossing) around $\hbar\omega \sim 0.23$ MeV. These are in good agreement with the values obtained 
using similar calculations for $^{193}$Tl by Reviol et al. \cite{22}. The same nomenclature as that 
of Ref. \cite{22} has been used in our calculations to label the quasiparticle levels. The observed band crossings
at $\hbar\omega \sim 0.28$ MeV and at $\hbar\omega \sim 0.37$ MeV in $^{193}$Tl were identified as due to the 
first and second neutron pair alignments by Reviol et al. The calculated crossing frequencies, however, are found to 
underpredict both the experimental crossing frequencies by the same amount and this discrepancy was attributed to 
the fact that the core deformation is not stiff as assumed in these calculations. In addition to the first proton
pair alignment, as in $^{193}$Tl, the first neutron pair alignment is also blocked in the odd-odd nucleus $^{194}$Tl. 
Therefore, the experimental band crossing in $^{194}$Tl, observed at a rotational frequency of $\hbar\omega \sim 0.34$ 
MeV, is due to the second neutron pair alignment. This value of the crossing frequency is in fairly good agreement with 
the observed second crossing in $^{193}$Tl but at a slightly lower frequency. The experimental crossing frequency in 
$^{194}$Tl is lowered by 0.03 MeV compared to the second crossing in $^{193}$Tl. This difference of the experimental crossing 
frequencies, in these two nuclei, is reproduced well in the calculated second neutron pair alignment frequencies for 
$^{194}$Tl ($\hbar\omega \sim 0.23$ MeV) obtained in our calculation, and that for the $^{193}$Tl 
($\hbar\omega \sim 0.26$ MeV) as obtained by Reviol et al. \cite{22}.

%%%%%%%%%%%%%%%%%%%%%%%%%%%%%%%%%%%%%%%%%%%%%%%%%%%%%%%%%%%
\begin{figure}[ht]
\begin{center}
\includegraphics*[scale=0.3,  angle =0]{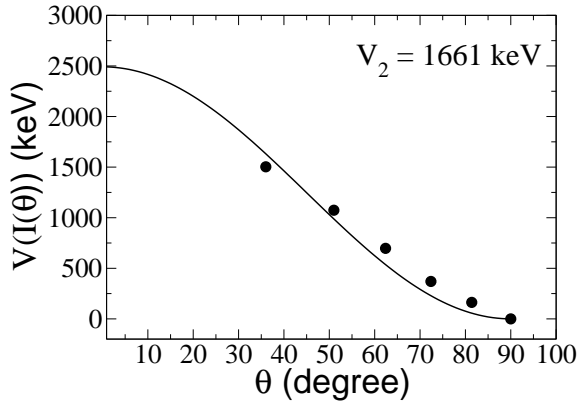}
\caption{The effective interaction between the angular momentum vectors, {\bf $j_\pi$} and {\bf $j_\nu$}, as a 
function of shears angle $\theta$ for the band B3 in $^{194}$Tl as obtained in the semiclassical formalism. }
\label{fig14}
\end{center}
\end{figure}
%%%%%%%%%%%%%%%%%%%%%%%%%%%%%%%%%%%%%%%%%%%%%%%%%%%%%%%%%%%

\subsection{Semiclassical calculations for the band B3}

To investigate the shears mechanism of the band B3, it has been studied by the semiclassical approach of 
Macchiavelli {\it et al.} \cite{15,23,24}. This is based on a schematic model of the coupling of two long j 
vectors, ($j_\pi$ and $j_\nu$), corresponding to the proton and the neutron parts of the angular momenta. This 
aims at extracting the information on the effective interaction between the nucleons which are involved in the 
shears mechanism. In this model, the shears angle ($\theta$), between the two j vectors, is an important variable 
which can be derived using the equation
\begin{equation}
cos\theta=\frac{{I}^2-{{j}_\pi}^2-{j_\nu}^2}{2\ {j_\pi}\ {j_\nu}}
\label{eqn:eqn3}
\end{equation}
where, $I$ is the total angular momentum. Considering the $\bm{j}_\pi$ and $\bm{j}_\nu$ values of 8.5$\hbar$ and 
13.5$\hbar$, respectively, for the proposed $\pi h^{2}_{9/2}s^{-1}_{1/2} \otimes \nu i^{-2}_{13/2} p_{3/2}$ configuration 
of the band B3, the band head spin is calculated to be 16 $\hbar$, assuming perpendicular coupling. This is in excellent 
agreement with the observed spin of the band head of this band. The maximum spin for this configuration has been 
calculated as $22^- \hbar$ corresponding to $\theta = 0^o$, which is again consistent with the highest spin observed 
for this band in the present work. Therefore, the angular momentum along the entire range of the band is, most likely, 
generated through the shears mechanism. The good agreement of the initial and the final spin values of this band with 
the calculated ones imply that the spins are generated solely by the shears mechanism with very little or no contribution 
from the rotation of the core. This fact is, again, corroborated well by the small quadrupole deformation obtained for 
this band in the TRS calculations.

According to the prescription of Macchiavelli {\it et al.} \cite{23}, the excitation energies of the states in shears
bands correspond to the change in the potential energy because of the recoupling of the angular momenta of the shears. 
The excitation energies of the states in the band, with respect to the band head energy, can be written as:
\begin{equation}
V(I(\theta)) = E_I - E_b = (3/2) V_2 \ cos^2{\theta}_I
\label{eqn:eqn4}
\end{equation}
where, $E_I$ is the energy of the level with angular momentum $I$, ${\theta}_I$ is the corresponding shears angle as
given in Eqn. 3, $E_b$ is the band head energy and $V_2$ is the strength of the interaction between the blades of 
the shears. Therefore, $V_2$ can be calculated by using the experimentally observed energy levels of the shears band. 
In Fig. 14, the $V(I(\theta))$ is plotted as a function of $\theta$. $V_2$ has been extracted from this plot by a fit of 
Eqn. 4. The fitted curve is shown as the solid line in Fig. 14. The extracted value of $V_2$ comes out to be 1661 
keV, which corresponds to an effective interaction of 332 keV per particle-hole pair for the suggested configuration
of the band B3.  This value of the effective interaction is in good agreement with the typical value of $\sim 300$ 
keV observed for the MR bands in Pb nuclei in this region \cite{clark01}.

\section{\bf Conclusion}
The $\gamma$-ray spectroscopy of the high spin states in the odd-odd nucleus $^{194}$Tl has been studied in the fusion evaporation 
reaction using $^{185,187}$Re target with $^{13}$C beam at 75 MeV. A new and improved level scheme of $^{194}$Tl is presented 
in this work which includes 19 new $\gamma$-ray transitions. The DCO ratio and the polarization asymmetry ratio measurements have 
been performed to assign the spins and parities of the levels. The $\pi h_{9/2} \otimes \nu i_{13/2}$ band (B1) in this nucleus has 
been extended just beyond the band crossing and up to the 18$^-$ $\hbar$ of spin. The uncertainties in the excitation energies and 
the spins in this band have been removed. The moment of inertia, alignment, the energy staggering and signature invesrion
of this band have been found to be very similar to its neighoring odd-odd isotopes. The TRS calculations predict similar 
structure for these nuclei and support the observed similarity. We have not found any indication of a chiral doublet band structure 
in $^{194}$Tl, unlike that which was reported for $^{198}$Tl. The observed band crossing in $^{194}$Tl could also be understood from
the band crossing in the neighboring odd-A nuclei with oblate deformation. Therefore, it needs further experimental and
theoretical investigation to understand the possible change in the structure of this 2-quaiparticle band in $^{198}$Tl, if any. 
Two new side bands (B2 and B3) have been observed in this work, for the {\it first time}, in $^{194}$Tl. 6-quasiparticle configurations 
for these bands have been suggested. A near spherical shape is predicted by the TRS calculations for the band B3 and is
consistent with the suggested MR nature of this band. This band has been discussed in the frame work of the semiclassical approach. 
The observed band head spin and the range of the spin values of this band are in agreement with such calculations. However, to get a 
microscopic understanding of this band, tilted axis cranking calculations are needed.\\

\section{\bf Acknowledgement}
The untiring effort of the operators and support staff at BARC-TIFR pelletron are acknowledged for providing a good beam 
of $^{13}$C. The authors gratefully acknowledge the efforts of Profs. R.G. Pillay and V. Nanal for the smooth running of 
the experiment and illuminating discussion. The help of the INGA community is gratefully acknowledged to set up the array 
and its associated electronics.


\begin{thebibliography}{99}
\bibitem{odd-Tl1} R.M. Diamond and F.S. Stephens, Nucl. Phys. {\bf A 45}, 632 (1963).
\bibitem{odd-Tl2} V.T. Gritsyna and H.H. Foster, Nucl. Phys. {\bf A 61}, 129 (1965).
\bibitem{odd-Tl3} J.O. Newton, S.D. Cirilov, F.S. Stephens and R.M. Diamond Nucl. Phys. {\bf A 148}, 593 (1970).
\bibitem{odd-Tl4} J.O. Newton, F.S. Stephens and R.M. Diamond Nucl. Phys. {\bf A 236}, 225 (1974).
\bibitem{1} R.M. Lieder et al., Nucl. Phys. {\bf A 299}, 255 (1978).
\bibitem{2} A.J. Kreiner et al., Phys. Rev. {\bf C 38}, 2674 (1988).
\bibitem{3} M.G. Porquet et al., Phys. Rev. {\bf C 44}, 2445 (1991).
\bibitem{4} W. Reviol et al., Phys. Scr. {\bf T 56}, 167 (1995).
\bibitem{5} I.G. Bearden et al., Nucl. Phys. {\bf A 576}, 441 (1994).
\bibitem{6} H. H\"{u}bel et al., Nucl. Phys. {\bf A 453}, 316 (1986).
\bibitem{7} C.Y. Xie et al., Phys. Rev. {\bf C 72}, 044302 (2005).
\bibitem{8} A.J. Kreiner et al., Phys. Rev. {\bf C 21}, 933 (1980).
\bibitem{9} A.J. Kreiner et al., Phys. Rev. {\bf C 20}, 2205 (1979).
\bibitem{10} A.J. Kreiner et al., Nucl. Phys. {\bf A 308}, 147 (1978).
\bibitem{11} A.J. Kreiner et al., Nucl. Phys. {\bf A 282}, 243 (1977).
\bibitem{12} A.J. Kreiner et al., Phys. Rev. {\bf C 23}, 748 (1981).
\bibitem{13} E.A. Lawrie et al., Phys. Rev. {\bf C 78}, 021305(R) (2008).
\bibitem{14} H. H\"{u}bel, Prog. Part. Nucl. Phys. {\bf 453}, 1 (2005).
\bibitem{15} R.M. Clark and A.O. Macchiavelli, Annu. Rev. Nucl. Part. Sci. {\bf 50}, 1 (2000).
\bibitem{aza90} F. Azaiez et al., Z. Phys. {\bf A 336}, 243 (1990).
\bibitem{aza91} F. Azaiez et al., Phys. Rev. Lett. {\bf 66}, 1030 (1991).
\bibitem{194Tl2009} P.L. Masiteng et al., Acta Phys. Pol. {\bf B 40}, 657 (2009).
\bibitem{ta08} H. Tan, et al., Nuclear Science Symposium Conference Record, NSS 08, IEEE, p 3196 (2008).
\bibitem{dsp1} R. Palit et al., Nucl. Inst. Meth. Phys. Res. {\bf A 680} 90 (2012)
\bibitem{dsp2} R. Palit, AIP Conf. Proc. {\bf 1336}, 573 (2011).
\bibitem{16} D. C. Radford, Nucl. Instrum. Methods Phys. Res. {\bf A 361}, 297 (1995).
\bibitem{17} A. Kr\"{a}mer-Flecken et al., Nucl. Instrum. Methods Phys. Res. {\bf A 275}, 333 (1989).
\bibitem{angcor} E.S. Macias et al., Computer Phys. Comm {\bf 11}, 75 (1976).
\bibitem{18} K. Starosta et al., Nucl. Instrum. Meth. Phys. Res. {\bf A 423}, 16 (1999).
\bibitem{19} Ch. Droste et al., Nucl. Instrum. Meth. Phys. Res. {\bf A 378}, 518 (1996).
\bibitem{ensdf} Balraj Singh, Nucl. Data Sheets {\bf 107}, 1531 (2006).
\bibitem{ensdf1} E. Achterberg, et al.,	Nucl. Data Sheets {\bf 107}, 1 (2006)	
\bibitem{22} W. Reviol et al., Nucl. Phys. {\bf A 548}, 331 (1992).
\bibitem{20} R. Bengtsson and S. Frauendorf, Nucl. Phys. {\bf A 327}, 139 (1979).
\bibitem{21} R. Bengtsson and S. Frauendorf, Nucl. Phys. {\bf A 314}, 27 (1979).
\bibitem{stag1} I. Hamamoto, Phys. Lett. {\bf B 235}, 221 (1990).
\bibitem{mrband} A.K. Jain et al., Pramana {\bf 75}, 51 (2010).
\bibitem{naza1} W. Nazarewicz, et al., Nucl. Phys. {\bf A 512} 61 (1990).
\bibitem{naza2} W. Nazarewicz, et al., Nucl. Phys. {\bf A 435} 397 (1985).
\bibitem{gm1} G. Mukherjee, et al., Phys. Rev. {\bf C 64} 034316 (2001).
\bibitem{gm2} G. Mukherjee, et al., Nucl. Phys. {\bf A 829} 137 (2009).
\bibitem{23} A.O. Macchiavelli {\it et al.}, Phys. Rev. C {\bf 57}, R1073 (1998).
\bibitem{24} A.O. Macchiavelli {\it et al.}, Phys. Rev. C {\bf 58}, R621 (1998).
\bibitem{clark01} R.M. Clark and A.O. Macchiavelli, Nucl. Phys. {\bf A682}, 415c (2001).


\end{thebibliography}
\end{document}